\documentclass[aps,prfluids,groupedaddress]{revtex4-2}

\usepackage{bm}
\usepackage{amsmath}
\usepackage{graphicx}
\usepackage{amsfonts}
\usepackage{mathtools} % for coloneq
\usepackage{newtxmath} % for mathscr
\usepackage{xcolor}    % for \color command
\usepackage{colortbl}  % for color tables

\begin{document}

%\preprint{}

% \title{Estimating the dimension of the chaotic attractor in} two-dimensional Kolmogorov flow}
\title{Latent-space variational data assimilation in two-dimensional turbulence}


\author{Andrew Cleary}
\email[]{acleary7@jhu.edu}
%\thanks{}
\affiliation{Department of Mechanical Engineering, Johns Hopkins University, Baltimore, MD 21218, USA}

\author{Qi Wang}
%\email[]{}
%\thanks{}
\affiliation{Department of Aerospace Engineering, San Diego State University, San Diego, CA 92182, USA}

\author{Tamer A. Zaki}
\email[]{t.zaki@jhu.edu}
%\thanks{}
\affiliation{Department of Mechanical Engineering, Johns Hopkins University, Baltimore, MD 21218, USA}
\affiliation{Department of Applied Math and Statistics, Johns Hopkins University, Baltimore, MD 21218, USA}


\date{\today}

\begin{abstract}
% currently ~230 words
% what is data assimilation
Starting from limited measurements of a turbulent flow, data assimilation (DA) attempts to estimate all the spatio-temporal scales of motion.
% Success is dependent on the observability of the system, namely how much of the initial turbulent field is encoded in the available subsequent measurements.
Success is dependent on whether the system is observable from the measurements, or how much of the initial turbulent field is encoded in the available measurements.
%Data assimilation (DA) refers to the infusion of experimental measurements of a turbulent flow into a numerical model of the flow.
% latent space data assimilation
Adjoint-variational DA minimizes the discrepancy between the true and estimated measurements by optimizing the initial velocity or vorticity field (the `state space').
Here we propose to instead optimize in a lower-dimensional latent space which is learned by implicit rank minimizing autoencoders.
% requirements for implementation
Assimilating in latent space, rather than state space, redefines the observability 
% of the system 
of the measurements
and identifies the physically meaningful perturbation directions which matter most for accurate prediction of the flow evolution. 
% testing on Kolmogorov flow
% order of magnitude improvement and scale scales
When observing coarse-grained measurements of two-dimensional Kolmogorov flow at moderate Reynolds numbers, the proposed latent-space DA approach estimates the full turbulent state with a relative error improvement of two orders of magnitude over the standard state-space DA approach.
% and one order of magnitude over state-space DA initialized with a super-resolution model.
The small scales of the estimated turbulent field are predicted more faithfully with latent-space DA, greatly reducing erroneous small-scale velocities typically introduced by state-space DA.
% noise
Furthermore, latent-space DA is demonstrated to be robust to noisy measurements at the range of Reynolds numbers considered.
% observability of the ground truth 
These findings demonstrate that the observability of the system from available data can be greatly improved when turbulent measurements are assimilated in the right space, or coordinates. 

\end{abstract}

%\keywords{}

\maketitle

\section{Introduction}
\label{sec:intro}

% short paragraph really motivating the approach
%DA is important.
Assimilating experimental data into numerical simulations improves the fidelity of the simulations and enables nonintrusive access to all the scales of the estimated flow.
%challenging problem, known to introduce artificial high wavenumber artefacts.
However, the estimation of turbulence from limited measurements is a difficult ill-posed problem \citep{zaki2025}.
Turbulence presents challenges such as the chaotic nature of the forward and dual problems, the non-uniqueness of solutions consistent with the measurements, and the introduction of erroneous small-scale velocities that decay over the solution trajectory.  
% motivate latent-space DA
% Here, we show that the accuracy of predicting turbulent trajectories from measurements can be significantly improved when assimilation is performed in a new space, as opposed to state space, which in the present work is a learned latent space. 
%Here, we show that the accuracy of predicting turbulent trajectories from measurements can be significantly improved when assimilation is performed, rather than in state space, in a new learned latent space. 

% turbulence paragraph
% In the study of turbulence, we rely on the interpretation of experimental measurements to probe the dynamics of the underlying flow.
% The conventional data-assimilation approach directly maps from the measurements to the turbulent state space. 
% We ask the question if our ability to observe the turbulence can be significantly improved by first mapping from the measurements to a pre-designed latent space, and subsequently to the full turbulent field?
% We demonstrate that this approach can lead to an accuracy improvement of two orders of magnitude over a much larger range of scales compared to existing techniques for the interpretation of turbulence measurements.

Conventional data assimilation utilizes the measurements to navigate the state-space representation of turbulence, and to directly estimate the velocity or vorticity field that justifies the measurements.
However, the state-space representation of turbulence is not necessarily suitable for this task. In this work, we propose to first map from the measurements to a pre-designed latent space, from which the full turbulent field can be subsequently decoded. We show that the accuracy of the estimated turbulent field can be significantly improved over a broad range of scales by interpreting the turbulent measurements in this latent space, when compared to the estimation using the state-space coordinates.

% paragraph on low-order models of turbulence, emphasize that they learn a more physically interpretable basis. 
The term latent space refers to a low-dimensional and interpretable representation of the turbulent field, whether by familiar modal decompositions \citep{taira2017modal}, or by non-linear autoencoder transformations \citep{brunton2020}.
Physical insights of complex dynamical systems can be gleaned from these low-dimensional latent spaces \citep{Fukami2023}.
%, and sparse system identification techniques can discover interpretable dynamical models in the associated co-ordinates \citep{champion2019}. 
%Low-dimensional modelling in terms of clusters computed via k nearest neighbors identified non-trivial quasi-attractors and transition processes \citep{Kaiser2014}.
Rank-minimizing autoencoders have successfully learned parsimonious representations of chaotic systems such as the Lorenz system, the Kuramoto-Sivashinsky equation and the lambda-omega reaction-diffusion system \citep{Zeng_2024}.
The latent spaces of such autoencoders have yielded insights on the nature of bursting events and the dynamical relevance of unstable periodic orbits in forced two-dimensional (2D) turbulence \citep{cleary2025}.

% data-driven approaches for inferring full state from  measurement, e.g. super-resolution
In the context of turbulence estimation, data-driven methods have been used to directly super-resolve limited instantaneous observations to the full flow state \citep[e.g.][]{Fukami_Fukagata_Taira_2019}.
Another super-resolution study considered a time-history of scarce measurements and inferred the pressure field of forced isotropic turbulent flow \citep{williams2024}.
% Only coarse data approaches
Machine-learning and classical adjoint DA methods to estimate turbulence have been pursued mostly separately, with a few exceptions including the following examples: 
% There is now interest in bridging classical DA and machine learning techniques to improve data assimilation.
\citet{DU2023} compared the estimation of wall turbulence using physics-informed neural networks and adjoint-variational techniques.  
\citet{Page_2025} modified the training of super-resolution networks to incorporate a time-forward Navier-Stokes evolution in the output, and a comparison to future data. Most recently, \citet{weyrauch2025} used the output of this super-resolution network as an initial guess to adjoint-variational DA. While this last effort has interfaced data-driven methods and adjoint-variational DA, the two techniques were not fully integrated.
% DA-inspired super-resolution network training protocols eliminate the need for access to fully-resolved turbulent fields and rely solely on measurement data \citep{Page_2025, weyrauch2025}.
% Similarly, operator networks can mitigate the computational expense of simulating high-speed boundary layer flows in DA problems \citep{CLARKDILEONI2023}.
%Once these networks have been adequately pre-trained, they can be directly used to assimilate sparse measurements in a `plug-and-play' mode.

Relatedly, some data-driven methods have been incorporated into statistical filtering approaches for data assimilation \citep{ozalp2026} and into adjoint-variational approaches to approximate the time forward operator \citep{CLARKDILEONI2023}, but neither of these cases exactly satisfy the governing dynamical equations.
In this work, we combine a learned space that is discovered using data-driven methods with the physics constraints of adjoint-variational DA.
This integration significantly enhances the observability of turbulence systems and their estimation across a wider range of scales, while still exactly satisfying the dynamical equations.
% Motivated by these recent advances, we introduce a powerful transformation such that measurements are assimilated in a pre-defined latent space, rather than in the state-space (discretized velocity or vorticity field) representation of the flow. The observability of the reference turbulent state from the measurements is greatly improved within this latent space, resulting in a more accurate reconstruction.
In \S\ref{sec:phys_da}, we outline the flow configuration studied and summarize the standard variational DA procedure. 
In \S\ref{sec:latent_da}, we present the proposed latent-space DA procedure. 
In \S\ref{sec:results}, we compare and interpret the improved performance of latent-space DA.
We conclude in \S \ref{sec:conclusion}. 

%\vspace*{-12pt}
\section{State-space data assimilation}
\label{sec:phys_da}

% \subsection{Flow configuration}
% \label{sec:flow_conf}

We consider Kolmogorov flow, which is monochromatically forced 2D turbulence on a square and doubly periodic domain \citep[][]{Chandler2013}. 
The out-of-plane vorticity $\omega = \partial_x v - \partial_y u$ defines the flow state in state space and evolves according to
\begin{equation}
    \partial_t \omega + \bm{u} \cdot \boldsymbol\nabla{\omega} = \frac{1}{Re} \nabla^2 \omega - k_f \cos {k_f y},
    \label{eq:kf_eq}
\end{equation}
where $\bm{u} = (u,v)$ is the velocity.
In this non-dimensionalization, the length scale $1/k^* = L^* / 2\pi$ is the inverse of the fundamental wavenumber (asterisk denotes dimensional quantities).
The time scale is $1/\sqrt{k^*\chi^*}$, where $\chi^*$ is the amplitude of the forcing in the momentum equation.  The Reynolds number is therefore $Re \coloneq \sqrt{\chi^*/k^{*3}}/\nu$, where $\nu$ is the kinematic viscosity.
The forcing wavenumber is set to $k_f = 4$.
Kolmogorov flow approaches an asymptotic regime beyond $Re \approx 50$ \citep{Cleary_Page_2025}, and therefore we consider $Re = \{40, 100, 400\}$ in this work. 
The vorticity-velocity Navier-Stokes equations (\ref{eq:kf_eq}) are solved using the pseudospectral version of the \texttt{JAX-CFD} solver \citep{Kochkov2021, LCspectral}, allowing for the efficient computation of gradients of the time-forward map of \eqref{eq:kf_eq} using automatic differentiation.
%At each time step in the solver, $\bm{u}$ is computed by solving the Poisson equation $\nabla^2 \psi = -\omega$, where the streamfunction $\psi$ is related to the induced velocity components via $u = \partial_y \psi, v = -\partial_x \psi$.
The computational grid is set to $N_x \times N_y = 128^2$ for $Re = \{40, 100\}$ and $512^2$ for $Re = 400$.

% \subsection{Variational data assimilation}
% \label{sec:4dvar}

\begin{figure*}
    \centering
    \includegraphics[width=0.8\linewidth]{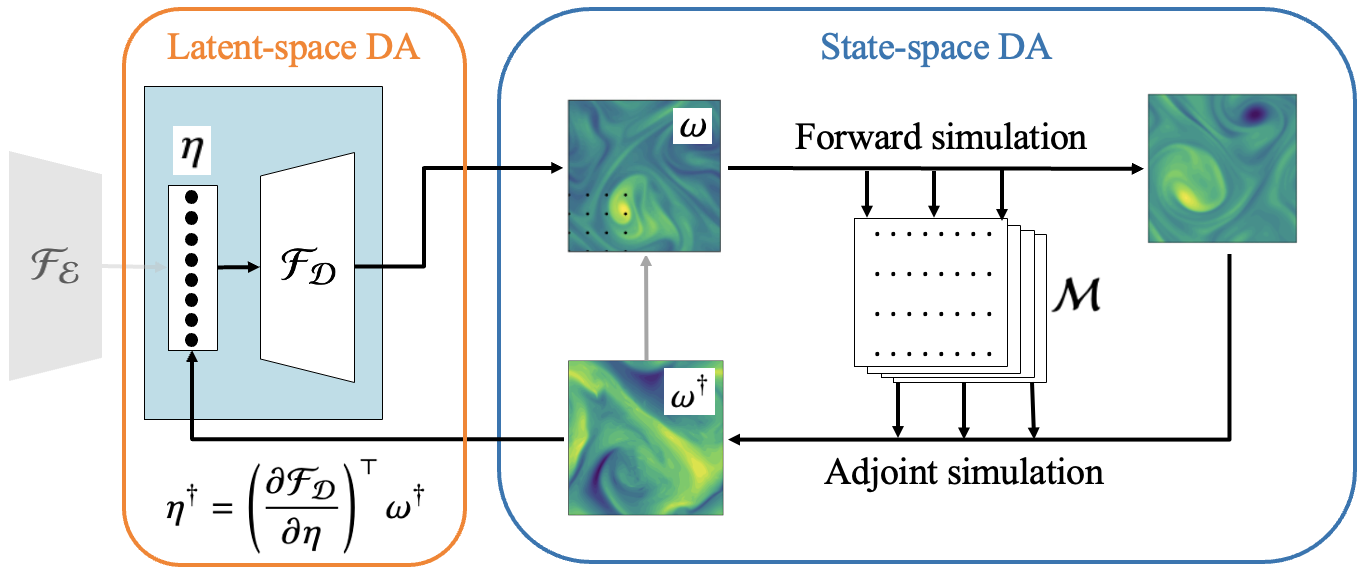}
    \caption{Schematic of latent-space data assimilation. The latent representation $\eta$ is mapped to state space $\omega$ by the decoder $\mathcal{F}_{\mathcal{D}}$, where the adjoint field $\omega^\dagger$ is computed. The gray arrow closes the standard, state-space `loop'. In latent-space assimilation, the latent state is updated by the transformed adjoint field $\eta^\dagger$. Sensor resolution is indicated by the lattice of black dots in the lower left corner of $\omega$.}
    \label{fig:latentDA}
%\vspace*{-6pt}
\end{figure*}

% sentence mentioning other DA approaches
%Many data assimilation approaches exist, such as linear stochastic estimation \citep{adrian1988stochastic, encinar2019logarithmic}, extended Kalman filtering \citep{chevalier2006state}, ensemble Kalman filtering \citep{colburn2011state} and ensemble-variational approaches \citep{mons2019kriging, mons2021ensemble}.
%In this paper, we adopt the variational method \citep{le1986variational, zaki2025}, which has been shown to estimate states which are more consistent with the measurements than both ensemble Kalman filtering and the ensemble-variational approach for two-dimensional unsteady flows around a cylinder \citep{mons2016reconstruction}.
Our objective is to estimate the flow state $\omega^*_0$ which, when evolved using the Navier-Stokes equations \eqref{eq:kf_eq}, reproduces available measurements $m_n^R = \mathcal{M}(\omega^R_n) \in \mathbb{R}^{d_m}$ from a reference solution $\omega^R$ at discrete times $t_n = n\Delta t$ over the time horizon $t_n \in [0, T]$ for $n = 0, \dots, N$.
The problem is formulated as a variational minimization of a cost function of the discrepancy between the estimated and true measurements,
%\vspace*{-6pt}
\begin{equation}
    \mathcal{J}(\omega_0) = \frac{1}{2} \sum_{n=0}^N \left\| \mathcal{M}(\omega_n) - m^R_n \right\|^2,
    %\vspace*{-6pt}
\label{eq:da_cost}
\end{equation}
subject to the constraint that $\omega_n = f^{t_n}(\omega_0)$ is the time-forward map of \eqref{eq:kf_eq} from initial condition $\omega_0$.
% highlight previous flow configurations with adjoint solvers
The required gradient of \eqref{eq:da_cost} with respect to $\omega_0$ can be computed using the discrete adjoint \citep{wang2019discrete} or automatic differentiation \citep{fan2025}.
%4DVar has been applied to the estimation of transitional and turbulent Taylor-Couette flow \citep{wang2019discrete}, turbulent channel flow \citep{foures2014data, wang2021state, wang2022observable}, two-dimensional Kolmogorov flow \citep{Page_2025} and to the reconstruction of small scales in three-dimensional Kolmogorov flow \citep{li2020small}.

% coarsening details
The measurement operator $\mathcal{M}$ is defined to be the coarse-graining operation $\mathcal{M}: \mathbb{R}^{N_x \times N_y} \to \mathbb{R}^{N_x/M \times N_y/M}$ which samples the high-resolution data at every $M^{\text{th}}$ gridpoint in both $x$- and $y$-directions.
The temporal coarsening $\Delta t = M\delta t$ is set by the same coarsening factor, where $\delta t$ is the time-step of the numerical simulation. 
At $Re = \{40, 100\}$, coarsening is set to $M=16$ and to $M=64$ at $Re = 400$, resulting in $8^2$ spatial measurements for each $Re$ case.
The DA time horizon is $T \approx 0.6 T_L$ where $T_L$ is the Lyapunov timescale at each $Re$.

As represented by the blue box in figure \ref{fig:latentDA}, the gradient of \eqref{eq:da_cost} can be computed by solving the adjoint equations,
\begin{subequations}
\begin{align}
\frac{\partial \omega^{\dagger}}{\partial \tau} + J(\psi, \omega^\dagger)  - \frac{1}{Re} \nabla^2 \omega^{\dagger} + \psi^{\dagger} &= \frac{\mathcal{D} \mathcal{J}}{\mathcal{D}\omega} \, , \\ % \label{eq:ad_omega}
\nabla^2 \psi^{\dagger} - J(\omega, \omega^\dagger) &= \frac{\mathcal{D} \mathcal{J}}{\mathcal{D}\psi} \, ,  %\label{eq:ad_psi} %\\
 %\frac{\partial \psi}{\partial x} \frac{\partial \omega}{\partial y} - \frac{\partial \psi}{\partial y} \frac{\partial \omega}{\partial x} &= J(\psi, \omega) \, , 
\end{align}
\label{eq:ad_both}%
\end{subequations}
backwards in time, where $\tau := T - t$, as derived in Appendix \ref{app:ad_deriv}.
The streamfunction $\psi$ is related to the velocity components via 
$u = \partial_y \psi, v = -\partial_x \psi$, 
and 
$J(\psi, \omega) = \frac{\partial \psi}{\partial x} \frac{\partial \omega}{\partial y} - \frac{\partial \psi}{\partial y} \frac{\partial \omega}{\partial x} $.
The adjoint field $\omega^{\dagger}$ at $t = 0$ yields the variation of the cost function with respect to the initial condition $\omega_0$,
\begin{equation}
    \frac{\mathscr D \mathcal J}{\mathscr D \omega_0} = \omega^{\dagger}(t = 0),
    \label{eq:GradJ}
\end{equation}
which can be used to update the estimated $\omega_0$ in a direction that minimizes the cost by better reproducing the measurements.  
In lieu of explicit solution of the adjoint equations \eqref{eq:ad_both}, we compute the gradient \eqref{eq:GradJ} by automatic differentiation.
% How to use gradient to optimize cost function
The cost function is then minimized using the Adam optimizer \citep{kingma2017adam} for a total of 500 optimization steps at each $Re$, which we found to be sufficient for the assimilation cost function to converge. 
A sweep over initial step sizes (or learning rates) was performed, such that $\alpha = 0.2$ was selected as the initial step size at each $Re$. 

Variational DA in state space requires a first guess of the initial flow field $\omega_0$. 
We will consider two initialization approaches. 
The first is a bicubic interpolation of the measurements (InterpDA).
The second is using a pre-trained super-resolution (SR) network $\mathcal{F}_{SR}: \mathbb{R}^{d_m} \to \mathbb{R}^{d_{\omega}}$, which maps from instantaneous measurements to instantaneous full resolution fields.
This initialization for data assimilation (SR-DA) was shown to yield significant improvement in the accuracy of the assimilation estimate by \citet{weyrauch2025}. 
The same network as in \citet{Page_2025} is adopted here, which is trained on the standard $L_2$ loss between the full resolution snapshot and the super-resolved snapshot. 
These SR networks were trained with the Adam optimizer with an initial learning rate of $10^{-4}$ and a batch size of 32, on data sets of 5000 turbulent snapshots at each $Re$ for a total of 250 epochs.
Full details on the data generation is given in Appendix \ref{app:data_irmae}.
% link to next section
These assimilations in state space serve as a benchmark for the proposed latent-space assimilation.

%\vspace*{-12pt}
\section{Latent-space data assimilation}
\label{sec:latent_da}

The classical variational-DA algorithm attempts to identify the optimal initial vorticity, or state-space representation of the flow, to reproduce the measurement.
The main idea of our latent-space DA is to, instead, attempt to predict a latent-space representation $\eta \in \mathbb{R}^{d_{\eta}}$ (see the orange box in figure \ref{fig:latentDA}).  
Shifting the object of the optimization to the latent space requires an evaluation of the gradient with respect to the new latent coordinates. 
The latent space is mapped to state space by a pre-trained decoder $\mathcal{F}_{\mathcal{D}}(\eta) = \omega \in \mathbb{R}^{d_{\omega}}$.
% give the DA cost function in latent space
The new DA cost function is then
\begin{equation}
    \mathcal{J}(\eta_0) = \frac{1}{2} \sum_{n=0}^N \left\| \left[\mathcal{M} \circ f^{t_n}\circ \mathcal{F}_{\mathcal D}\right](\eta_0) - m^R_n \right\|^2,
    \label{eq:latent_cost}
\end{equation}
where the latent state estimate $\eta_0$ is first decoded to state space, then \eqref{eq:kf_eq} is solved in state space and finally the model observations are compared to available measurements (combined orange and blue boxes in figure \ref{fig:latentDA}).
% give the gradient transformation between spaces
The adjoints in latent and state space are related by the Jacobian 
% of the decoder 
$\partial\mathcal{F}_{\mathcal{D}} / \partial \eta \in \mathbb{R}^{d_{\omega}\times d_{\eta}}$,
\begin{equation}
    \eta^\dagger = \left(\frac{\partial\mathcal{F}_{\mathcal{D}}}{\partial \eta}\right)^\top \omega^\dagger.
    \label{eq:latent_adj}
\end{equation}
% As such, latent space assimilation can be easily coupled to existing adjoint solvers in any flow configuration, as long as the vector-Jacobian product of the decoder can be computed. 
% Alternatively, as the numerical solver of \eqref{eq:kf_eq} used here is fully-differentiable, we directly compute $\eta^{\dagger}$ from \eqref{eq:latent_cost}.  
% benefits of time marching in state space
We note that no dynamical model in the latent space is required and all time marching is performed in state space, such that the estimated solution $f^{t}(\mathcal{F}_{\mathcal D}(\eta_0))$ for $t\in[0,T]$ exactly satisfies the Navier-Stokes equations \eqref{eq:kf_eq}.
Furthermore, given a pre-existing adjoint solver for the flow in question, only the ability to compute vector-Jacobian products of the decoder is required to perform latent-space DA according to \eqref{eq:latent_adj}.
In this case, the vector-Jacobian product can be computed efficiently by a single reverse-mode automatic differentiation of the decoder, such that the full Jacobian does not need to be computed.
In practice, we built the decoder $\mathcal{F}_{\mathcal{D}}$ using Keras with the JAX backend, allowing us to evaluate and differentiate \eqref{eq:latent_cost} as a single computational graph.
As in state-space DA, the Adam optimizer is used to minimize \eqref{eq:latent_cost}, and the same number of optimization steps were taken. 
A sweep over initial step sizes led to the choices $\alpha = \{5, 0.2, 0.02\}$ at $Re = \{40,100, 400\}$. 

\subsection{The latent space}
\label{sec:irmae}

% introduce IRMAE
We adopt the latent space of the implicit rank-minimizing autoencoder (IRMAE). 
Unlike standard autoencoders, IRMAE is distinguished by the addition of a series of fully-connected, linear layers in the bottleneck of the network, which minimize the rank of the covariance matrix of the latent representation \citep{Zeng_2024}.
This effect, known as implicit rank minimization, has been demonstrated empirically when optimizing linear multi-layer neural networks with gradient descent \citep{jing2020}.
The addition of these linear layers into the bottleneck of the network does not increase the expressivity of the network, as they can be represented by a single matrix multiplication.
% motivate why IRMAE
The IRMAE architecture has demonstrated comparable performance to variational autoencoders for smooth interpolation in the latent space and generating new samples from random noise \citep{jing2020}.
As such, the latent representation exhibits these favorable properties in addition to being approximately minimal rank.
%For a detailed description of the IRMAE architecture, 
% training hyperparameters,  reconstruction errors, 
%and characterization of the latent space, see \citet{cleary2025}. 

% Training
The IRMAE network $\mathscr{A}$ seeks to learn the identity function
\begin{equation}
    \mathscr{A}(\omega) \equiv [\mathcal{F}_{\mathcal{D}} \circ \mathcal{W} \circ \mathcal{F}_{\mathcal{E}}](\omega) \approx \omega,
    \label{eq:id_irmae}
\end{equation}
where the encoder $\mathcal{F}_{\mathcal{E}} : \mathbb{R}^{d_{\omega}} \to \mathbb{R}^{d_{\eta}}$ maps the input vorticity snapshot to a low-dimensional representation ($d_{\eta} = 1024$), $\mathcal{W} : \mathbb{R}^{d_{\eta}} \to \mathbb{R}^{d_{\eta}} $ represents a series of four fully-connected, equally-sized linear layers (pure matrix multiplication) within the embedding space, and the decoder $\mathcal{F}_{\mathcal{D}}$ is defined as above.
The encoder and decoder consist of a series of convolutional dense blocks at varying resolutions. 
%, as in \citet{cleary2025}. 

To train the networks at $Re = \{40, 100, 400\}$, datasets were generated by sampling long-time trajectories of the flow at every time unit, resulting in datasets with a total number of snapshots $N_S \approx \{6, 10, 10\} \times 10^4$, respectively.
Initial transient periods of 50 time units were discarded from the trajectories, such that the training data is sampled from the turbulent attractor (for full details, see Appendix \ref{app:data_irmae}).
The networks are then trained to minimize the loss function
\begin{equation}
    \mathscr{L} = \frac{1}{N_S} \sum_{j=1}^{N_S} \left\| \mathscr{A}(\omega^j) - \omega^j \right\|^2,
    \label{eq:irmae_loss}
\end{equation}
where each $\omega^j$ is a full-field snapshot from the training dataset.
Full details of the IRMAE architecture, training protocol and reconstruction accuracy of \eqref{eq:id_irmae} are provided in Appendix \ref{app:irmae_arch}.
For each $Re$ considered, DA was performed on independent trajectories which were never seen during the training of the neural networks.

% \subsection{Initialization of flow estimate}
% \label{sec:init}

% \begin{figure}
%     \centering
%     \includegraphics[width=0.8\linewidth]{Figures/InitDA.png}
%     \caption{Schematic of the initialization of the data assimilation in state space $\omega_0$ and in latent space $\eta_0$.}
%     \label{fig:da_init}
% \end{figure}

Variational DA in latent space requires a first guess of the initial latent representation $\eta_0$. 
This initialization is obtained by using the aforementioned pre-trained SR network followed by the IRMAE encoder, i.e., $\eta_0 = \mathcal{F}_{\mathcal E} \circ \mathcal{F}_{SR}(m^R_0)$. 
Comparable performance can be achieved by training a fully-connected network to map directly from measurement space to the IRMAE latent space.

\section{Results}
\label{sec:results}

\subsection{Accuracy of the estimated fields}
\label{sec:accuracy}

\begin{figure*}
    \centering
    \includegraphics[width=0.9\linewidth]{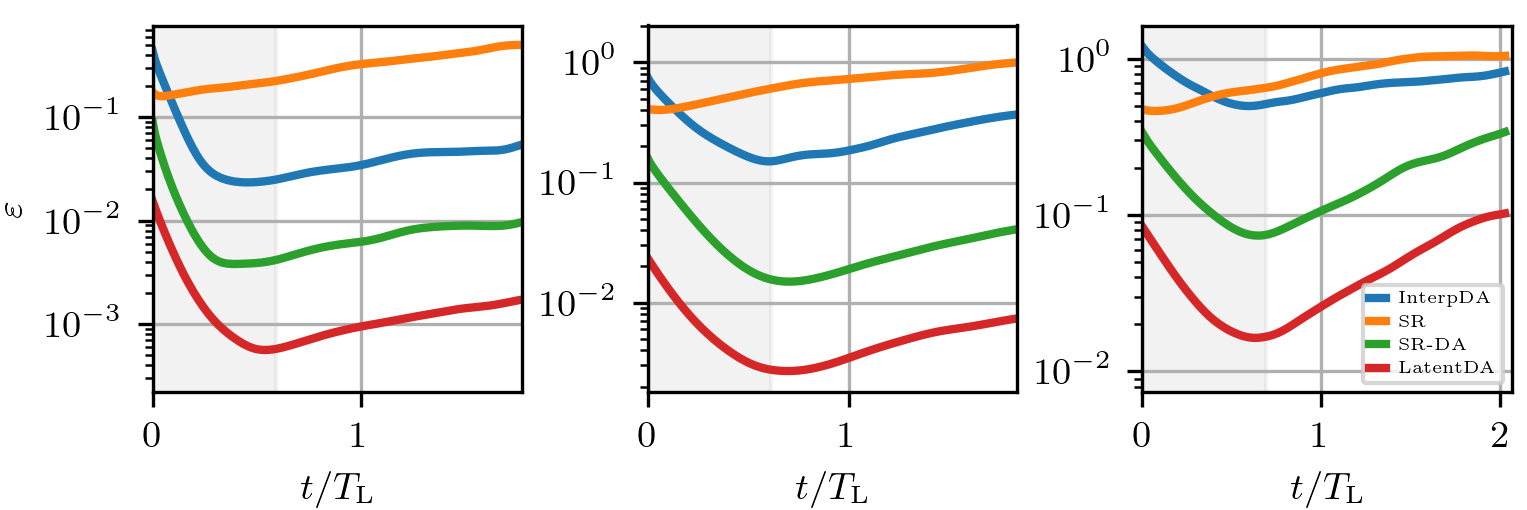}
    \caption{Relative error $\varepsilon$ of the estimated turbulent field by (blue) InterpDA, (orange) SR, (green) SR-DA and (red) LatentDA at (left) $Re = 40$, (middle) $Re = 100$ and (right) $Re = 400$ as a function of time normalized by the Lyapunov timescale. Bold colored lines: ensemble mean of ten independent trajectories. Gray region marks the data-assimilation time horizon.}
    \label{fig:Re_all}
\end{figure*}

The accuracy of the flow trajectories is compared when the initial flow states are estimated from a single inference of the super-resolution network (SR), super-resolution-initialized DA in state space (SR-DA), and DA in latent space (LatentDA).
Standard state-space DA initialized with a bicubic interpolation (InterpDA) is presented as a benchmark. 
The evolution of the error,
% \begin{equation}
%     \varepsilon(t) = \frac{\| f^t(\omega^*_0) - f^t(\omega^R_0) \|}{\| f^t(\omega^R_0) \|}
% \end{equation}
\begin{equation}
    \varepsilon(t) = {\| f^t(\omega^*_0) - f^t(\omega^R_0) \|}\,/\,{\| f^t(\omega^R_0) \|},
\end{equation}
for each method at $Re = \{40, 100, 400\}$ is reported in figure \ref{fig:Re_all}, where $\omega^*_0$ and $\omega^R_0$ are the estimated and reference turbulent fields at $t=0$. 
The spatio-temporal coarsening factor was $M = 16$ at both $Re = \{40, 100\}$ and $M = 64$ at $Re = 400$, and DA was performed over the time horizon $T \approx 0.6 T_L$ which is indicated by the gray shaded region in figure \ref{fig:Re_all}.
The Lyapunov timescales, $T_L = \{ 8.5, 3.3, 2.9 \}$ for $Re = \{40, 100, 400\}$ respectively, were computed using the method by \citet{benettin1976}.

% \begin{figure}
%     \centering
%     \includegraphics[width=\linewidth]{Figures/Re100_initialstates.png}
%     \caption{Comparison of (a) reference $\omega^R_0$ at $t=0$ and the estimated field $\omega^*_0$ from (i-iv) InterpDA, SR, SR-DA, LatentDA at $Re = 100$. (b) Contours of the out-of-plane vorticity, (c) enstrophy spectra $\Omega(k)$ as a function of wavenumber $k$. The black spectra and the black contours in (biv) denote the reference data.}
%     \label{fig:Re100_omega0}
% \end{figure}

\begin{figure*}
    \centering
    \includegraphics[width=\linewidth]{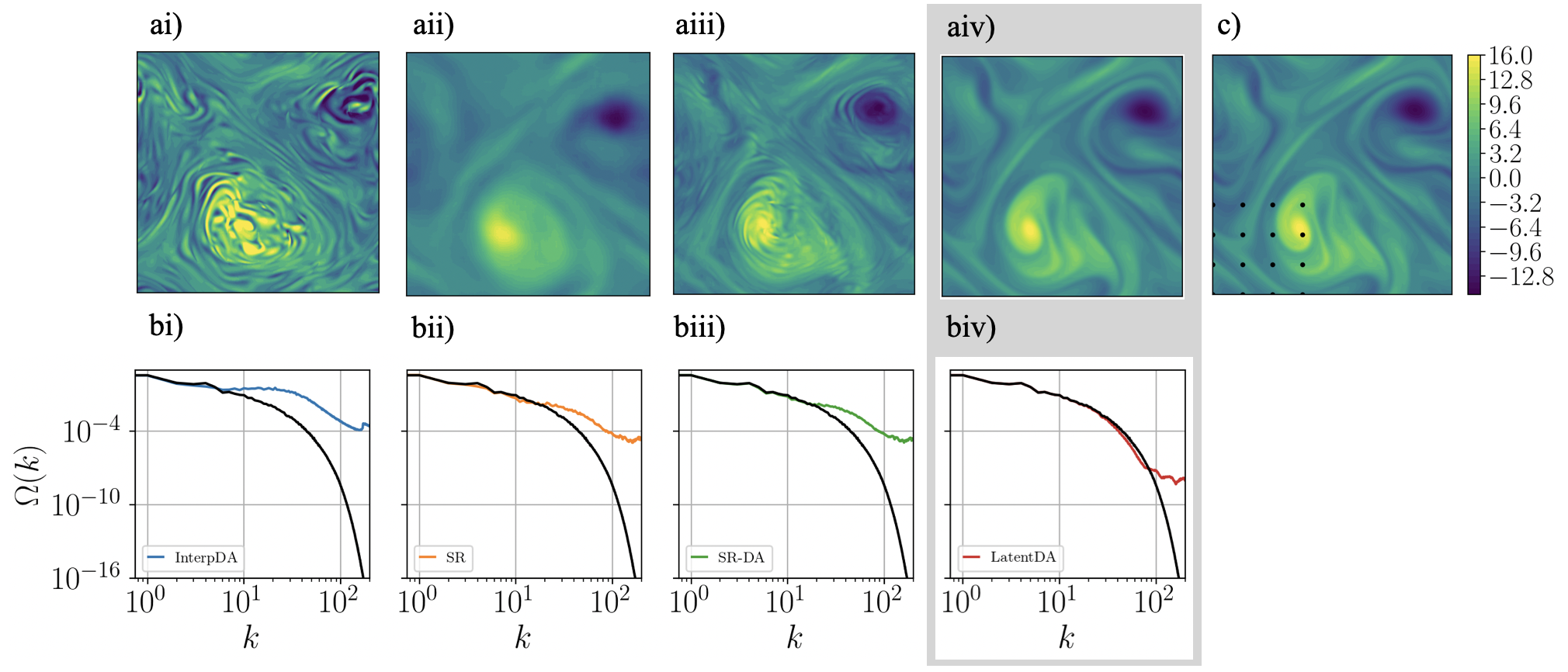}
    \caption{Comparison of the estimated field $\omega^*_0$ from (i-iv) InterpDA, SR, SR-DA and LatentDA (highlighted by gray box) with (c) reference $\omega^R_0$ at $t=0$ at $Re = 400$. The lattice of black dots in (c) indicates the sensor resolution. 
    % (a) Contours of the out-of-plane vorticity, and 
    (b) Enstrophy spectra $\Omega(k)$ as a function of wavenumber $k$. The black spectra denote the reference data. }
    \label{fig:Re400_omega0}
\end{figure*}

% \begin{table}[]
%     \centering
%     \begin{tabular}{c|ccc>{\columncolor[gray]{0.9}}c}
%         $Re$ & InterpDA & SR & SR-DA & LatentDA \\ \hline
%         40 &  0.025 $\pm$ 0.017 & 0.224 $\pm$ 0.065 &  0.004 $\pm$ 0.008  &  0.0006 $\pm$ 0.0004 \\
%         100 &  0.150 $\pm$ 0.192 &  0.607 $\pm$ 0.044  &  0.016 $\pm$ 0.003  & 0.0028 $\pm$ 0.0004 \\
%         400 & 0.516 $\pm$ 0.258  &  0.655 $\pm$ 0.068  & 0.075 $\pm$ 0.042   & 0.017 $\pm$ 0.0110 \\
%     \end{tabular}
%     \caption{Mean and standard deviation of the relative error $\varepsilon$ of the estimated turbulent field at the DA time horizon $t = T$, over an ensemble of ten independent trajectories.}
%     \label{tab:latentda}
% \end{table}

\begin{table}[]
    \centering
    \begin{tabular}{c|ccc>{\columncolor[gray]{0.9}}c}
        $Re$ & InterpDA & SR & SR-DA & LatentDA \\ \hline
        ~40 ~&  ~~~2.5 $\pm$ ~1.7 & ~~22.4 $\pm$ 6.5 &  ~~0.4 $\pm$ 0.8  &  ~~0.06 $\pm$ 0.04~~ \\
        100 ~&  ~~15.0 $\pm$ 19.2 &  ~~60.7 $\pm$ 4.4  &  ~~1.6 $\pm$ 0.3  & ~~0.28 $\pm$ 0.04~~ \\
        400 ~& ~~51.6 $\pm$ 25.8  &  ~~65.5 $\pm$ 6.8  & ~~7.5 $\pm$ 4.2   & ~~1.70 $\pm$ 1.10~~ \\
    \end{tabular}
    \caption{Mean and standard deviation of the percent relative error ($\varepsilon \times 100$) of the estimated turbulent field at the DA time horizon $t = T$, over an ensemble of ten independent trajectories.}
    \label{tab:latentda}
\end{table}

LatentDA (red) results in an improvement of approximately two orders of magnitude over the standard InterpDA approach (blue) and one order of magnitude over SR-DA (green)  at $Re = \{ 40 ,100 \}$.
Even at $Re = 400$, we still observe an improvement by more than one order of magnitude over the InterpDA approach, and a significant reduction of ${\sim}50\%$ error over the best-performing approach in state space (SR-DA).
Furthermore, table \ref{tab:latentda} reveals that the standard deviation of the percent relative error at the DA time horizon $T$ using LatentDA is significantly smaller than SR-DA at $Re = 400$. 
The decrease in $\varepsilon(t)$ for InterpDA, SR-DA and LatentDA up to $t = T$ is typical of variational DA: the error between model predictions and measurements late in the assimilation time horizon has the largest impact on the gradient direction as these perturbations can amplify under the chaotic nature of the adjoint equations \eqref{eq:ad_both} \citep{zaki2021}.
In contrast, the SR (orange) estimate only takes the measurements at $t = 0$ into account, such that its error increases over the assimilation time horizon.
%
% Not only is the estimate of the initial state $\omega^*_0$ much improved using LatentDA, but the growth rate of $\varepsilon(t)$ after $t = T$ is consistent across all three DA approaches considered, such that predictions can be made over much longer time intervals.
Since the accuracy advantage of the estimated trajectories from LatentDA is retained throughout the observation horizon, and because the growth rate of $\varepsilon(t)$ after $t = T$ is consistent across all three DA approaches, accurate predictions from LatentDA can be made over much longer time intervals.
For example, at $Re = 100$ and using LatentDA, the prediction error is ${\sim}1\%$ at $2.5T_L$ (not shown), which is a four folds longer horizon than when using SR-DA. 

% To simplify the exposition, only results from the SR networks trained on the standard MSE loss \eqref{eq:sr_mse} are presented here and throughout.
% Although SR networks trained on the coarse dynamics loss function \eqref{eq:sr_tc} exhibited comparable values of $\varepsilon(t)$ after a short time window, they exhibited larger $\varepsilon(0)$ than the MSE networks, consistent with the results in \citet{Page_2025}.
% This can be attributed to a poorer reconstruction of the small spatial scales -- these high wavenumber artefacts quickly decay under the action of viscosity. 
%These artefacts are less apparent in the $\omega^*_0$ fields estimated by the MSE networks as these models are trained using the full resolution training data. 

A comparison of the estimated $\omega^*_0$ fields at $Re = 400$ using each approach is presented in figure \ref{fig:Re400_omega0}.
When the data is assimilated in state space (figure \ref{fig:Re400_omega0}i), $\omega^*_0$ exhibits unphysical high-wavenumber artifacts and does not at all resemble the reference data (figure \ref{fig:Re400_omega0}c). 
In contrast, the estimated field by SR (figure \ref{fig:Re400_omega0}ii) is overly smooth, which is a symptom of the spectral bias of neural networks.  
State-space DA initialized with this field (SR-DA) introduces some high-frequency errors (figure \ref{fig:Re400_omega0}iii), but the enstrophy spectrum $\Omega(k)$ remains poorly representative of the true field.  
When the assimilation is performed in latent space (figure \ref{fig:Re400_omega0}iv, highlighted by the gray box), the predicted initial field is most accurate and $\Omega(k)$ is most consistent with the reference turbulent state.
Similar trends were observed at the lower $Re$ considered.
%These high-wavenumber artefacts are a less-discussed feature of variational data assimilation, and warrant a brief discussion. 

\begin{figure*}
    \centering
    \includegraphics[trim=0 5 0 0, clip, width=\linewidth]{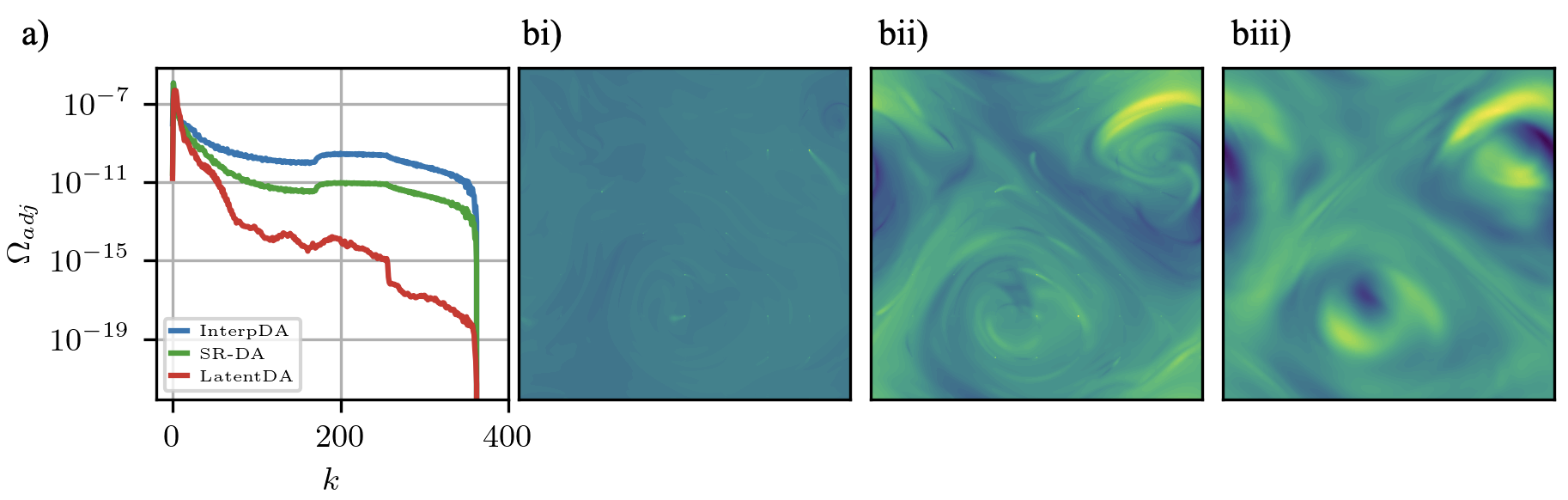}
    \caption{(a) The spectra $\Omega_{adj}$ of the initial normalized adjoint field in state space $\omega^{\dagger}$ for (blue) InterpDA, (green) SR-DA and (red) LatentDA adjoint field decoded to state space $\mathcal{F}_{\mathcal D}(\eta + \alpha \eta^{\dagger}) - \mathcal{F}_{\mathcal D}(\eta)$ at $Re = 400$. (b) Visualization of these adjoint fields in state space for (bi) InterpDA, (bii) SR-DA and (biii) LatentDA.}
    \label{fig:adjoint_Re400}
\end{figure*}

To understand the cause of the high-wavenumber artifacts in state-space DA, it should be noted that the adjoint equations \eqref{eq:ad_both} are forced by $\mathcal{D}\mathcal{J} / \mathcal{D}\omega$, which is a series of singular impulses in space and time at each measurement location. 
These superpositions of delta functions are advected and diffused by \eqref{eq:ad_both}, but signatures of this forcing are very apparent in the visualization of $\omega^{\dagger}(t=0)$ at $Re = 400$ in figures \ref{fig:adjoint_Re400}(bi,bii).
As the Fourier transform of a delta function is a constant function of $k$, these singular impulses lead to high wavenumber spectral content of the state-space adjoint variables. % (figure \ref{fig:adjoint_Re400}a). % in figure \ref{fig:adjoint_Re400}.
Each variational DA iteration slightly perturbs the turbulent field estimate in a direction with high energy in the high wavenumbers. %, resulting in the high wavenumber artefacts evident in figure \ref{fig:Re100_omega0}.
For latent-DA, the effective adjoint update can be visualized in state space by evaluating % decoding the updated field and then subtracting the non-updated field 
$\mathcal{F}_{\mathcal D}(\eta + \alpha \eta^{\dagger}) - \mathcal{F}_{\mathcal D}(\eta)$ (figure \ref{fig:adjoint_Re400}biii), which does not contain these singular forcing signatures. 
% , where $\alpha$ is the step size used in the optimization. 
To enable comparison between the state-space and latent-space adjoint spectra, the adjoint directions are normalized before their spectra are computed.
As shown in figure \ref{fig:adjoint_Re400}a, the relative energy in the high wavenumbers of this effective latent adjoint direction is many orders of magnitude smaller than the state-space adjoint directions, which is consistent with the improved spectrum of $\omega_0^*$ for LatentDA. % in figure \ref{fig:Re100_omega0}.

\subsection{Robustness to noisy measurements}

\begin{figure}
    \centering
    \includegraphics[width=0.8\linewidth]{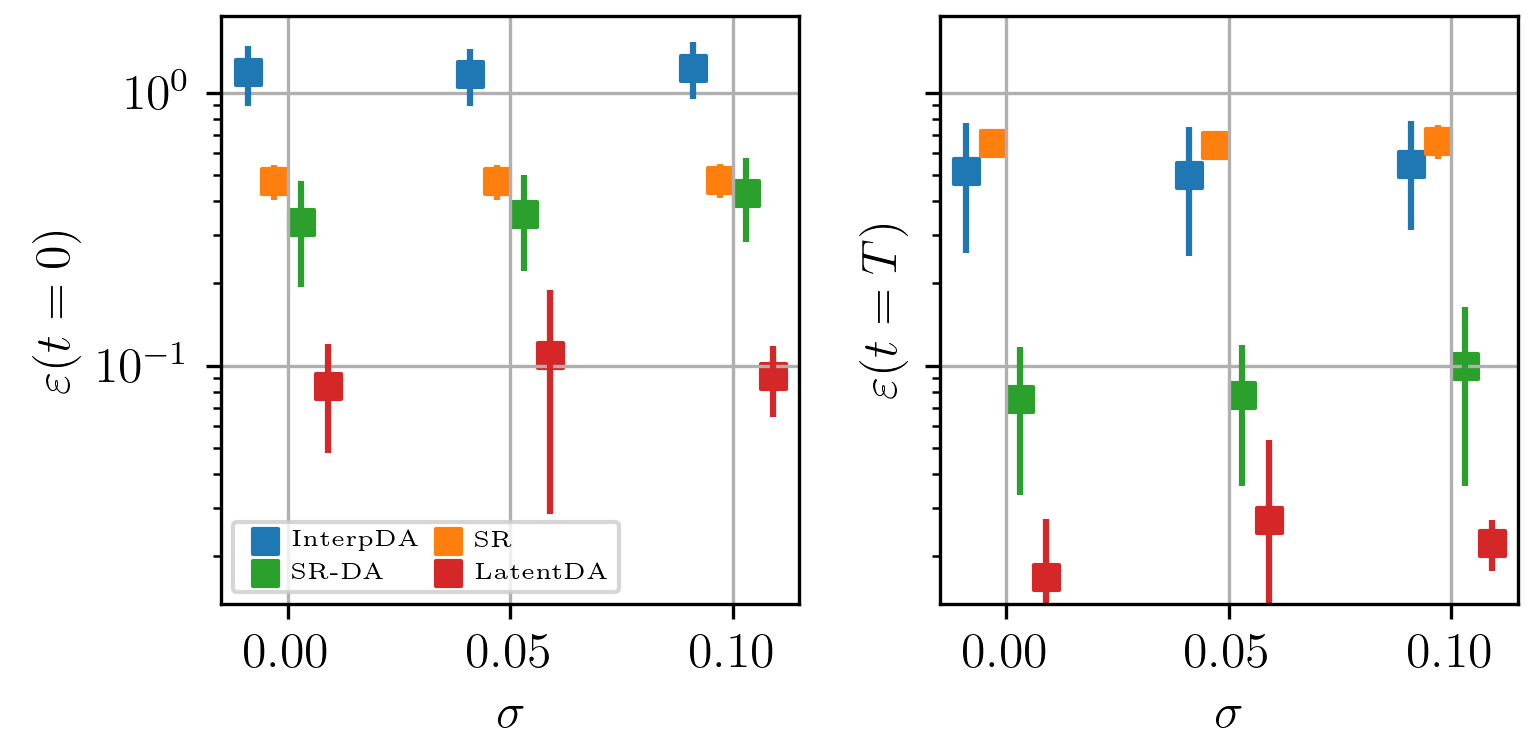}
    \caption{Relative error of the estimated turbulent field at (left) the initial time $\varepsilon(t = 0)$ and (right) the DA time horizon $\varepsilon(t = T)$ as a function of the noise amplitude $\sigma$, by (blue) InterpDA, (orange) SR, (green) SR-DA and (red) LatentDA at $Re = 400$ (markers are slightly offset for clarity).}
    \label{fig:noise_error}
\end{figure}

In practice, the measurement data available for assimilation are typically not perfect, but rather corrupted by some sensor error. 
We briefly investigate the effect of measurement noise on LatentDA as a test of the robustness of the proposed method. 
Similarly to \citet{wang2021state} and \citet{Page_2025}, noise is added to the measurements of the reference trajectories
\begin{equation}
    m^R_{n,j} \gets m^R_{n,j} + \Sigma_{n,j},
\end{equation}
where $m^R_n = \{ m^R_{n,j} \}_{j=1}^{N_xN_y/M^2}$ are the measurements at each discrete time $t_n$, and the noise is normally distributed based on the local absolute vorticity $\Sigma_{n,j} \sim \mathcal{N}(0, \sigma |m^R_{n,j}|)$.
We consider noise amplitudes of $\sigma \in \{ 0, 0.05, 0.1 \}$, and the same noise is also present when initializing each variational procedure, whether by interpolation (InterpDA) or by a pre-trained SR network inference (SR, SR-DA and LatentDA). 

The relative error of the estimated turbulent state at the initial time $t = 0$ and at the DA time horizon $t= T$ are reported as a function of $\sigma$ in figure \ref{fig:noise_error} for $Re = 400$. 
All methods are similarly robust to the increasing noise at both times considered. 
As such, LatentDA maintains its reconstruction accuracy advantage over all state-space assimilation methods when measurements are noisy.
This is a promising result in the context of scaling LatentDA to higher-$Re$, experimental measurements.

\subsection{Observability in latent space}
\label{sec:discussion}

We consider two dynamical perspectives that explain the improved performance of LatentDA. 
The first perspective recalls that the IRMAE models have been trained on data sampled from the turbulent attractor $\omega \in \mathcal{A}$ and have learned latent representations $\eta$ such that $\mathcal{F}_{\mathcal{D}}(\eta) \in \mathcal{A}$ holds approximately. 
Let us consider the gradient direction in latent space, $\alpha\eta^{\dagger}$, as a small perturbation such that $\mathcal{F}_{\mathcal{D}}(\eta - \alpha \eta^{\dagger}) \in \mathcal{A}$ still holds approximately.
%This is not an unreasonable assumption given the maximum magnitude attained by enstrophy spectrum of the effective decoded adjoint field in figure \ref{fig:adjoint_Re400}.
By expanding this small perturbation to linear order,
\begin{align*}
    \mathcal{F}_{\mathcal{D}}(\eta - \alpha \eta^{\dagger}) &= \mathcal{F}_{\mathcal{D}}(\eta) - \alpha\frac{\partial \mathcal{F}_{\mathcal D}}{\partial \eta}\eta^{\dagger} + \dots \in \mathcal{A},
\end{align*}
it is clear that the columns of the decoder Jacobian are the perturbation directions which, to linear order, approximately remain on the turbulent attractor. 
The linear transformation of adjoints from state to latent space \eqref{eq:latent_adj} can now be understood as a projection onto these physically relevant perturbation directions. The associated linearized update in state space is given by,
\begin{align*}
    \omega_0^* \approx \mathcal{F}_{\mathcal{D}}(\eta) - \alpha\frac{\partial \mathcal{F}_{\mathcal D}}{\partial \eta} \left(\frac{\partial \mathcal{F}_{\mathcal D}}{\partial \eta} \right)^{\top} \omega^{\dagger}.
\end{align*}
As such, %the estimate remains physically relevant in state space throughout the latent variational DA method, to linear order. %, explaining the lack of high wavenumber artefacts in figures \ref{fig:Re100_omega0} and \ref{fig:Re400_unroll}.
the methodological interpretation of this linearization is that the decoding process ensures that the estimate remains approximately on the attractor in state space.
Physically, these improved perturbation directions are marked by an absence of the high-wavenumber artifacts, as shown in figure \ref{fig:adjoint_Re400}.
In contrast, a traditional state-space adjoint update visibly perturbs some of the dimensions of the estimate away from the turbulence attractor by the addition of erroneous energy in the high wavenumbers.

The second perspective considers how observable the reference initial turbulent field is from the measurements in state or latent space.
This perspective naturally leads to the idea of the observability Gramian and its eigendecomposition, which we will compute at optimality and relate to the local geometry of the DA cost function.
%As we will show here, the iterative gradient-based updates throughout the variational DA method can be expressed as expansions in some basis of adjoint fields.
When assimilating data variationally, it is beneficial for the reference turbulent field $\omega^R_0$ to be well represented by the basis spanned by these eigenvectors, such that the iterative updates can effectively converge onto $\omega^R_0$.
%\AC{This basis will be defined by the eigenvectors of the Hessian of the DA cost function, which measures the local curvature of the loss landscape.}
In order to derive an analytic expression for the observability Gramian, we begin by defining the deviation field $w = \omega - \omega^R$ which is governed by the linearized Navier-Stokes equations 
\begin{equation}
    \partial_t w - J(\varphi, \omega^R) - J(\psi^R, w) = \frac{1}{Re}\nabla^2w,
    \label{eq:ns_lin}
\end{equation}
with the associated deviation streamfunction $\varphi = -\nabla^2 w$ and the ground truth streamfunction $\psi^R = -\nabla^2 \omega^R$.
As shown in \citet{wang2022observable} for turbulent channel flow, the DA cost function \eqref{eq:da_cost} can be written in terms of the measurement kernel $\phi(\bm x_m)$ which extracts the measurement of interest at location $\bm x_m$
\begin{equation}
    \mathcal{J}(w_0) = \frac{1}{2} \sum_{n=0}^N \sum_{m=1}^{d_m} \left\langle w(t=t_n), \phi(\bm x_m) \right\rangle^2,
    \label{eq:da_cost_dev1}
\end{equation}
where $\langle a,b \rangle = \int_{V} ab \, dV$ is the spatial inner product over the computational domain $V$. 
In this study $\phi(\bm x_m) = \delta(\bm x - \bm x_m)$ is the Kronecker delta function, but more complex kernels can be defined as required. 
Forward-adjoint duality can be exploited to write \eqref{eq:da_cost_dev1} in terms of the adjoint field $w^{\ddagger}$, as
\begin{align*}
    \left\langle w(t=t_n), \phi(\bm x_m) \right\rangle &= \left\langle \mathcal{L} w_0, \phi(\bm x_m) \right\rangle \\ &= \left\langle  w_0, \mathcal{L}^{\ddagger}\phi(\bm x_m) \right\rangle \\  &= \left\langle  w_0, w^{\ddagger}(t=0;t=t_n,\bm x_m) \right\rangle,
\end{align*}
where $\mathcal{L}$ is the forward operator of the linearized Navier-Stokes equations \eqref{eq:ns_lin} which advances $w_0$ to $w(t=t_n)$.
Note that $w^\ddagger \neq \omega^\dagger$, and an explicit relation between them will be given below.
The associated adjoint operator $\mathcal{L}^\ddagger$ solves the adjoint of the linearized equations
\begin{subequations}
    \begin{align}
    \frac{\partial w^{\ddagger}}{\partial \tau} + J(\psi^R, w^{\ddagger})  - \frac{1}{Re} \nabla^2 w^{\ddagger} + \varphi^{\ddagger} &= 0 \, , \\ %\label{eq:ad_lin_omega} \\
    \nabla^2 \varphi^{\ddagger} - J(\omega^R, w^{\ddagger})  &= 0 \, , %\label{eq:ad_lin_psi}
    \end{align}
    \label{eq:ad_lin_both}%
\end{subequations}
with the initial condition $w^{\ddagger}(\tau = 0) = \phi(\bm x_m)$. 
This duality can be used to rewrite \eqref{eq:da_cost_dev1} as
\begin{equation}
    \mathcal{J}(w_0) = \frac{1}{2} \sum_{n=0}^N \sum_{m=1}^{d_m} \left\langle w_0, w^\ddagger(t = 0; t=t_n, \bm x_m) \right\rangle^2.
    \label{eq:da_cost_dev2}
\end{equation}
Expressing the DA cost function in terms of the deviation field $w$ yields the following explicit expression of the gradient of the DA cost function at optimality,
\begin{align}
    \omega^\dagger(t=0)\bigg |_{\omega = \omega^R} = \frac{\mathscr D \mathcal J}{\mathscr D w_0} \bigg |_{w_0 = 0} = \sum_{n=0}^N \sum_{m=1}^{d_m} \left\langle w_0, w^\ddagger(t = 0; t=t_n, \bm x_m) \right\rangle w^\ddagger(t = 0; t=t_n, \bm x_m) = 0 \,,
    \label{eq:da_cost_grad}
\end{align}
and its Hessian, % \emph{at} optimality},
\begin{align}
    \mathcal H \coloneq \frac{\mathscr D^2 \mathcal J}{\mathscr D w_0 \mathscr D w_0} \bigg |_{w_0 = 0}= \sum_{n=0}^N \sum_{m=1}^{d_m} w^\ddagger(t = 0; t=t_n, \bm x_m) w^\ddagger(t = 0; t=t_n, \bm x_m) \, ,
    \label{eq:da_cost_hess}
\end{align}
% The gradient NEAR optimality (15) is a weighted superposition of the principal directions of curvature AT optimality. 
% If the expression for the gradient holds near optimality, then the expression for the Hessian should hold near optimality also (not just at optimality)?
%The gradient at optimality \eqref{eq:da_cost_grad} is an expansion in the basis spanned by $w^\ddagger(t=0;t=t_n,\bm x_m)$, while
which is the covariance matrix of the adjoint fields of the linearized equations $w^{\ddagger}$.
The eigendecomposition $\mathcal{H}v_j = \lambda_j v_j$ with eigenvalues $\lambda_i$ and eigenvectors $v_j$ then
establishes the principal directions of local curvature at optimality.
As $\mathcal{H}$ is the observability Gramian \citep{rowley2005model}, the leading $v_j$ also indicate the directions in state space to which the measurements are most sensitive.
It is beneficial for the reference state $\omega^R_0$ to have a strong projection on these modes as they are the most observable directions.
%\AC{As the gradient update lives in the subspace spanned by this basis, it can only introduce or remove physical structures which have a nonzero projection on this basis.
%For this reason, the interpretability of the measurements can be greatly improved if the ground truth has a strong projection on this basis.}
%The Hessian  can be written as the cross-correlation $\mathcal H = A A^\top$ of the matrix $A$ with these adjoint fields as columns.

\begin{figure*}
    \centering
    \includegraphics[width=0.9\linewidth]{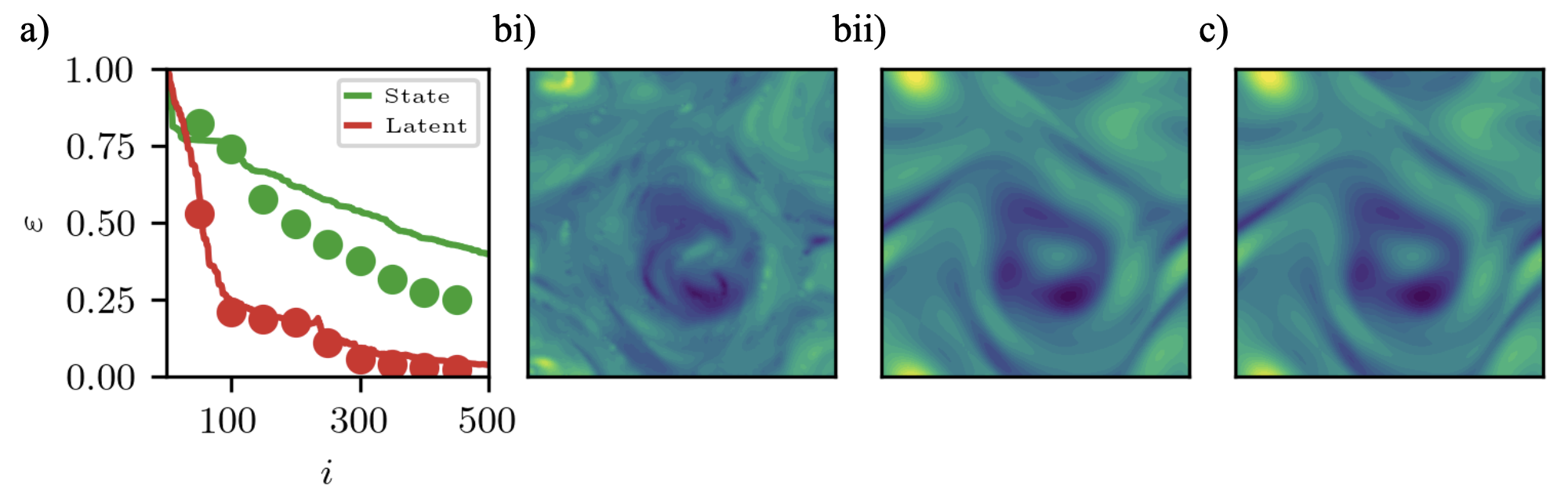}
    %\vspace*{-4pt}
    \caption{(a) The relative error $\varepsilon$ of the instantaneous reconstruction of the reference turbulent field $\omega^R_{0,i}$ (lines) and the time-averaged relative error over the DA time horizon (circles) using the leading $i$ observability modes in (green) state and (red) latent space at $Re = 100$. (b) Contours of the out-of-plane vorticity of the reconstructed field with 500 observability modes in (i) state and (ii) latent space, and (c) the reference turbulent field.}
    \label{fig:proj_onto_hessian}
\end{figure*}

%a proper orthogonal decomposition (POD) basis 
% \citep{taira2017modal} 
%for these adjoint fields. 
As such, we examine how well $\omega^R_0$ can be reconstructed by the leading $v_j$ at $t = 0$, 
\begin{equation}
    \omega^R_{0,i} = \sum_{j=0}^i \left\langle v_j, \omega^R_0 \right\rangle v_j.
    \label{eq:ref_recon_adj_basis}
\end{equation}
The relative error $\varepsilon$ of the reconstruction is reported in figure \ref{fig:proj_onto_hessian} for $Re = 100$, where we compare SR-DA (green) and $\eta_0^*$ obtained with LatentDA (red).
In state space, $\mathcal{H}$ is computed at the reference $\omega^R_0$, while in latent space $\mathcal{H} = \mathscr D^2 \mathcal J / \mathscr D \eta_0 \mathscr D \eta_0$ is computed at the encoded reference state $\mathcal{F}_{\mathcal{E}}(\eta^R_0)$.
The first 500 adjoint POD modes were computed by the Arnoldi iteration in state and latent space, such that the full Hessian matrix was not required to be computed explicitly. 
In practice, as both the decoder $\mathcal{F}_{\mathcal{D}}$ and the flow solver were built using the JAX backend, the Jacobian-vector product function in JAX was chained with the gradient function to return the Hessian-vector product in either state or latent space, as part of the Arnoldi iteration.
The reconstructions were also time-evolved over the DA time horizon, and the time-averaged relative errors (circles) were computed as a function of the number of adjoint POD modes used.  
The reference turbulent state can be reconstructed to ${\sim} 5\%$ relative error in the latent adjoint basis with 500 adjoint POD modes, as opposed to ${\sim} 40\%$ relative error in the state-space adjoint basis, demonstrating the improved observability of $\omega^R_0$ from the measurements in the latent space near optimality.
Perhaps more notably, ${\sim} 100$ modes in latent space can reconstruct the reference turbulent state with ${\sim} 25\%$ error, while the same number of modes can only reconstruct the state with ${\sim} 75\%$ error in state space.
High wavenumber artifacts and signatures of the localized adjoint forcing are evident in the state-space adjoint basis reconstruction (figure \ref{fig:proj_onto_hessian}bi). These small-scale fluctuations decay over the DA time horizon, explaining the improved time-averaged relative errors in state space.  In contrast, the decoded reconstruction in latent space (figure 6bii) very closely resembles the reference turbulent state (figure 6c), and the error in the representation of the initial state and the time-averaged errors during the evolution are similar.  These results demonstrate that the observability of the turbulence using limited measurements can be appreciably improved when the assimilation is performed in the right space, or coordinates.

\section{Conclusion}
\label{sec:conclusion}

% \item Recap latent space DA.
In the study of turbulence, we often rely on the interpretation of measurements to probe the dynamics of the underlying flow.  Data assimilation seeks to map from the measurements to the associated turbulent state that satisfies the Navier-Stokes equations. We ask the question if our ability to observe the turbulence can be significantly improved by first mapping from the measurements to a pre-designed latent space, and subsequently to the full turbulent field? We demonstrate that the mapping to the latent intermediate coordinates, namely the latent space of a pre-trained autoencoder, can lead to significant accuracy improvement in the interpretation of turbulence measurements.

% Specifically, we introduced a new variational data assimilation approach that, rather than operate in state space, exploits the improved observability in the latent space of a pre-trained autoencoder. 
%When a chaotic system is perturbed, the projection of this perturbation on the leading Lyapunov direction will amplify fastest.
In state-space DA, the adjoint field is highly localized in the vicinity of the delta-function forcing, such that the spectrum of the adjoint field is broadband. 
The updates in latent space are more targeted and physically relevant, resulting in more accurate reconstruction of both the large and small scales, and a smaller departure from the truth over time.
%Although the small-scale velocities that are introduced by these broadband perturbation directions dissipate under the action of viscosity, ...
%
% \item Emphasise order of magnitude improvement.
An improvement of an order of magnitude was achieved when assimilating data in latent space at $Re = 40$ and 100, compared to the best approach considered in state space.  
A lower but nonetheless appreciable improvement was achieved at $Re = 400$, where the increased spatiotemporal complexity of the flow requires a larger latent space dimensionality.
The small scales of the estimated flow state are more dynamically relevant at all $Re$ considered.
LatentDA also exhibited a comparable robustness to noisy measurements as all state-space DA methods.
The training of the IRMAE models is a once-off offline computation. The models at $Re = 40$ and $Re = 100$ comprised $6.2\times 10^6$ trainable parameters, and training for 500 epochs on an NVIDIA A100 (80GB) GPU required approximately two days.  The larger model at $Re = 400$ comprised $8.5 \times 10^6$ trainable parameters. The training data in this case exceeded the available memory on the GPU, and training required four weeks. Once trained, the stark improvement in assimilation accuracy is achieved with a negligible increase in run-time computational expense over classical state-space variational data assimilation.
Assimilation in latent space requires an extra projection of the state-space adjoint into latent co-ordinates \eqref{eq:latent_adj}, which can be computed by a single reverse-mode backpropagation through the pre-trained decoder.
Relative to the cost of state-space DA, this extra projection step increases the computational time of each adjoint loop by just 1\%.
This small fraction is due to the low cost of a single backpropagation through the decoder, in contrast to the large computational cost of the traditional state-space forward-adjoint loop (equivalent to two direct numerical simulations).
We anticipate that the relatively humble computational overhead of the latent projection step will become yet more negligible when scaling to three-dimensional flows.

% \item Next steps: Investigate if we can scale to fully turbulent 3D flows like subdomains of  channel flow.
This work demonstrates the potential benefits of combining variational and data-driven techniques to interpret turbulence measurements.  The observability of turbulence from the data is much improved by taking advantage of the latent space of the autoencoder.  More generally, the results demonstrate the power of exploiting such new latent spaces in the study of turbulence.  
% In this work, coarsened measurements of 2D Kolmogorov flow were assimilated.  Future work could consider more spatiotemporally complex flows. % , or more challenging available measurement data such as scarce and non-uniformly separated sensors. 
%Shallow recurrent decoders \citep{williams2024} may provide suitable initializations in the latent space in such situations. 
% We hope to report on the performance of latent space DA in such configurations in the near future.

\begin{acknowledgments}

This work was supported by the Air Force Office of Scientific Research (AFOSR) grants FA9550-21-1-0345 and FA9550-25-1-0011.

\end{acknowledgments}

\appendix

\section{Continuous adjoint equations: vorticity-streamfunction formulation}
\label{app:ad_deriv}

The 2D velocity-vorticity Navier-Stokes equations \eqref{eq:kf_eq} with some forcing $g$ can be written in terms of the vorticity and streamfunction $\psi$ as
\begin{equation}
    \frac{\partial \omega}{\partial t} - J(\psi, \omega) - \frac{1}{Re}\nabla^2 \omega + g = 0,
    \label{eq:kf_vortstream}
\end{equation}
where 
\begin{equation}
    J(\psi, \omega) = \frac{\partial \psi}{\partial x} \frac{\partial \omega}{\partial y} - \frac{\partial \psi}{\partial y} \frac{\partial \omega}{\partial x}
    \label{eq:poissonbracket}
\end{equation}
and $\psi$ is related to the velocity components via $u = \partial_y \psi, v = -\partial_x \psi$, such that $\psi$ is recovered from $\omega$ by solving the Poisson equation $\nabla^2 \psi = -\omega$.
The constrained optimization problem \eqref{eq:da_cost} can be cast as an unconstrained optimization problem by defining the Lagrangian
\begin{equation}
    \mathscr{L} = \mathcal{J} - \left( \omega^{\dagger}, \frac{\partial \omega}{\partial t} - J(\psi, \omega) - \frac{1}{Re}\nabla^2\omega + g \right) - \left( \psi^{\dagger}, \omega + \nabla^2\psi \right),
    \label{eq:lagrangian}
\end{equation}
where the inner product is defined by integrating over space $V$ and the observation time horizon $t \in [0, T]$
\begin{equation}
    \left( a,b \right) = \int_0^T \int_{V} ab \, \, dV dt .
\end{equation}
Taking variations of $\mathscr{L}$ with respect to the Lagrange multipliers $\omega^{\dagger}$ and $\psi^{\dagger}$ yields the Navier-Stokes equation for the estimated $\omega$ and the Poisson equation relating $\omega$ to $\psi$. 
To take variations of $\mathscr{L}$ with respect to $\omega$ and $\psi$, we proceed as follows.
Assuming periodic boundary conditions, one can re-express terms such as $\left( \omega^{\dagger}, J(\psi, \omega) \right)$ by integrating by parts in space
\begin{align*}
    \left( \omega^{\dagger}, J(\psi, \omega) \right) &= \int_0^T \int_{V} \omega^{\dagger} \left[ \frac{\partial \psi}{\partial x} \frac{\partial \omega}{\partial y} - \frac{\partial \psi}{\partial y} \frac{\partial \omega}{\partial x} \right] \, \, dV dt, \\
    &= \iint \left[ \frac{\partial}{\partial x} \left(\omega^{\dagger} \frac{\partial \psi}{\partial y} \right) \omega - \frac{\partial}{\partial y}\left(\omega^{\dagger} \frac{\partial \psi}{\partial x} \right) \omega \right] dV dt, \\
    &= -\left( J(\psi, \omega^{\dagger}), \omega \right),
\end{align*}
where in the final equality, we assume commutativity of the partial derivatives. 
By the antisymmetry of $J$, the relation $\left( \omega^{\dagger}, J(\psi, \omega) \right) = \left( J(\omega, \omega^{\dagger}), \psi \right)$ follows quickly. 
By isolating $\omega$ in the remaining terms, the Lagrangian $\eqref{eq:lagrangian}$ can be written as
\begin{align}
    \mathscr{L} = \mathcal{J} -& \int_V\left(\omega^{\dagger}(T)\omega(T) + \omega^{\dagger}(0)\omega_0\right)dV  \nonumber \\ &+  \left( \frac{\partial \omega^{\dagger}}{\partial t} - J(\psi, \omega^{\dagger}) + \frac{1}{Re} \nabla^2 \omega^{\dagger} - \psi^{\dagger} , \omega \right) \nonumber \\ &- \left(  \omega^{\dagger}, g \right) - \left(  \nabla^2 \psi^{\dagger}, \psi \right) .
\end{align}
Taking variations of $\mathscr{L}$ with respect to $\omega$ is now straightforward, and variations with respect to $\psi$ follow quickly from the antisymmetry of $J$.
Setting $\mathcal{D}\mathscr{L} / \mathcal{D}\omega = 0$ and $\mathcal{D}\mathscr{L} / \mathcal{D}\psi = 0$ yields the equations for the adjoint fields $\omega^{\dagger}$ and $\psi^{\dagger}$ 
\begin{align}
    \frac{\partial \omega^{\dagger}}{\partial \tau} + J(\psi, \omega^{\dagger}) - \frac{1}{Re} \nabla^2 \omega^{\dagger} + \psi^{\dagger} &= \frac{\mathcal{D} \mathcal{J}}{\mathcal{D}\omega}, \\
    \nabla^2 \psi^{\dagger} - J(\omega, \omega^{\dagger}) &= \frac{\mathcal{D} \mathcal{J}}{\mathcal{D}\psi},
\end{align}
where $\tau = T - t$ is the backwards time. 
Then, the adjoint field at $t = 0$ yields the total variation of the cost function with respect to the initial condition $\omega_0$,
\begin{equation}
    \frac{\mathscr D \mathcal J}{\mathscr D \omega_0} = \omega^{\dagger}(t = 0).  %\frac{\mathcal D \mathscr L}{\mathcal D \omega_0} =
\end{equation}

\section{Data Generation}
\label{app:data_irmae}

Datasets were generated at each $Re$ by sampling long trajectories of the flow.
At $Re = \{ 40, 100\}$, 100 trajectories of length $10^3$ were sampled at every advective time unit.
Initial transient periods of 50 time units were discarded from the trajectories, such that the training data is sampled from the turbulent attractor.
The resulting dataset was resampled according to the dissipation rate of each snapshot, as in \citet{cleary2025}. 

At $Re = 400$, four trajectories of length $2.5\times 10^4$ were sampled at every time unit, resulting in $N_S = 10^5$ snapshots.
Again, initial transient periods of 50 time units were discarded from these trajectories.
These snapshots were not resampled according to dissipation rate, as i) it is more computationally expensive to generate large datasets at the increased $Re$ and increased resolution of $512^2$ and ii) the dissipation rate distribution is less skewed at the higher $Re$.

\section{IRMAE architecture}
\label{app:irmae_arch}

\newcommand{\pc}[2]{\text{PC}(#1 \times #1, #2)}
\newcommand{\lpc}[2]{\text{LPC}(#1 \times #1, #2)}
\newcommand{\db}[1]{\text{DB}(#1 \times #1)}
\newcommand{\p}[1]{\text{MP}(#1 \times #1)}
\newcommand{\up}[1]{\text{UP}(#1 \times #1)}

\subsection{$Re = 40$ \& $100$}

The IRMAE architecture used at $Re = 40$ and $100$ is adopted from \citet{cleary2025}. 
Dense blocks (DB) \citep{Huang2016} are groups of convolutional layers where the output of each layer is concatenated with its input.
Here, each dense block is made up of three convolutional layers, each with 32 filters.
We apply the `GELU' activation function \citep{hendrycks2023} and batch normalization \citep{ioffe15} to the output of convolutional layers. 
The encoder architecture is given by the following sequence of operations:
\begin{align}
    \omega &\to \pc{8}{32} \to \db{8} \to \p{2} \nonumber\\
    &\to \pc{4}{32} \to \db{4} \to \p{2} \nonumber\\
    &\to \pc{4}{32} \to \db{4} \to \p{2}  \nonumber\\
    &\to \pc{4}{32} \to \db{4} \nonumber \\
    &\to \pc{4}{4} \to \text{Flatten} = \mathcal{F}_{\mathcal{E}}(\omega),
\end{align}
where `PC' stands for a convolutional layer with periodic padding and `MP' for a max pooling layer. 
The first term in the brackets denotes the size of the convolutional filters and the second term denotes the number of filters.

The bottleneck $\mathcal W$ consists of a series of four fully connected (`FC') linear layers with equal input/output dimension,
\begin{equation}
    \mathcal{F}_{\mathcal{E}}(\omega) \to FC(1024)^4 = \mathcal{W}(\mathcal{F}_{\mathcal{E}}(\omega)).
\end{equation}

The latent space is the output of this bottleneck, which is then reshaped into an image with four channels of dimension $16^2$.
The decoder then mirrors the encoder, with the addition of an output stream of linear periodic convolutional layers (`LPC'):
\begin{align}
    \mathcal{W}(\mathcal{F}_{\mathcal{E}}(\omega)) &\to \text{Reshape} \nonumber\\
    &\to \pc{4}{32} \to \db{4} \to \up{2} \nonumber\\
    &\to \pc{4}{32} \to \db{4} \to \up{2} \nonumber\\
    &\to \pc{4}{32} \to \db{4} \to \up{2}  \nonumber\\
    &\to \pc{8}{32} \to \db{8} \nonumber \\
    &\to \lpc{16}{1} \to \lpc{12}{1} \nonumber\\
    &\to \lpc{8}{1} \to \lpc{6}{1} \nonumber\\ 
    &= \mathcal{F}_{\mathcal{D}}(\mathcal{W}(\mathcal{F}_{\mathcal{E}}(\omega))) = \mathscr{A}(\omega),
\end{align}
where `UP' denotes upsampling layers. 

The $Re = 40$ and $100$ networks were initialized with the final network weights from \citet{cleary2025}, which were trained on symmetry-reduced snapshots (both continuous and discrete symmetries of Kolmogorov flow).
We refer the reader to \citet{cleary2025} for full details on the training hyperparameters, symmetry reduction protocol and reconstruction error of these network weights.

However, we do not assume that the measurements are taken from symmetry-reduced turbulent states here, so these network weights are further fine-tuned on data which is augmented at every epoch with random symmetry operations.
The AdamW optimizer \cite{loshchilov2019} is used for this transfer learning, with an initial learning rate of $10^{-4}$ and the default weight decay coefficient of 0.004 for 500 epochs.

\subsection{$Re = 400$}

The architecture of IRMAE at $Re = 400$ was slightly modified from that used at $Re = 40$ and $100$, such that the input and output vorticity fields would be of resolution $512 \times 512$.
One extra downsampling/upsampling operation was inserted into the encoder/decoder, such that the input field was reduced to have dimension $32^2$, rather than $16^2$.
The kernel sizes were also modified throughout the encoder and decoder to reflect the increased dimensionality of the input/output vorticity fields, as outlined below. 

The encoder architecture is given by the following sequence of operations:
\begin{align}
    \omega &\to \pc{8}{32} \to \db{8} \to \p{2} \nonumber\\
    &\to \pc{8}{32} \to \db{8} \to \p{2} \nonumber\\
    &\to \pc{8}{32} \to \db{8} \to \p{2} \nonumber\\
    &\to \pc{4}{32} \to \db{4} \to \p{2}  \nonumber\\
    &\to \pc{4}{32} \to \db{4} \nonumber \\
    &\to \pc{4}{1} \to \text{Flatten} = \mathcal{F}_{\mathcal{E}}(\omega),
\end{align}
The input vorticity field is a single-channel image of size $512^2$, hence the output of the encoder $\mathcal{F}_{\mathcal{E}}(\omega)$ is a flattened vector of length $32^2 = 1024$. 

The architecture of the bottleneck is unchanged from the lower $Re$ bottleneck architecture.
The latent representation is then reshaped to have dimension $32^2$.
The decoder architecture is again slightly modified to mirror the encoder, 
\begin{align}
    \mathcal{W}(\mathcal{F}_{\mathcal{E}}(\omega)) &\to \text{Reshape} \nonumber\\
    &\to \pc{4}{32} \to \db{4} \to \up{2} \nonumber\\
    &\to \pc{4}{32} \to \db{4} \to \up{2} \nonumber\\
    &\to \pc{8}{32} \to \db{8} \to \up{2}  \nonumber\\
    &\to \pc{8}{32} \to \db{8} \to \up{2}  \nonumber\\
    &\to \pc{8}{32} \to \db{8} \nonumber \\
    &\to \lpc{16}{1} \to \lpc{12}{1} \nonumber\\
    &\to \lpc{8}{1} \to \lpc{6}{1} \nonumber\\ 
    &= \mathcal{F}_{\mathcal{D}}(\mathcal{W}(\mathcal{F}_{\mathcal{E}}(\omega))) = \mathscr{A}(\omega).
\end{align}

% Re = 400 difference
The $Re = 400$ IRMAE network was trained from scratch to input/output states on the finer computational grid of $512^2$.
The AdamW optimizer is used with an initial learning rate of $10^{-4}$ and the default weight decay coefficient of 0.004 for 400 epochs.
The training data is symmetry augmented at every epoch with random symmetry operations.

\subsection{Reconstruction accuracy}

\begin{figure}
    \centering
    \includegraphics[width=0.8\linewidth]{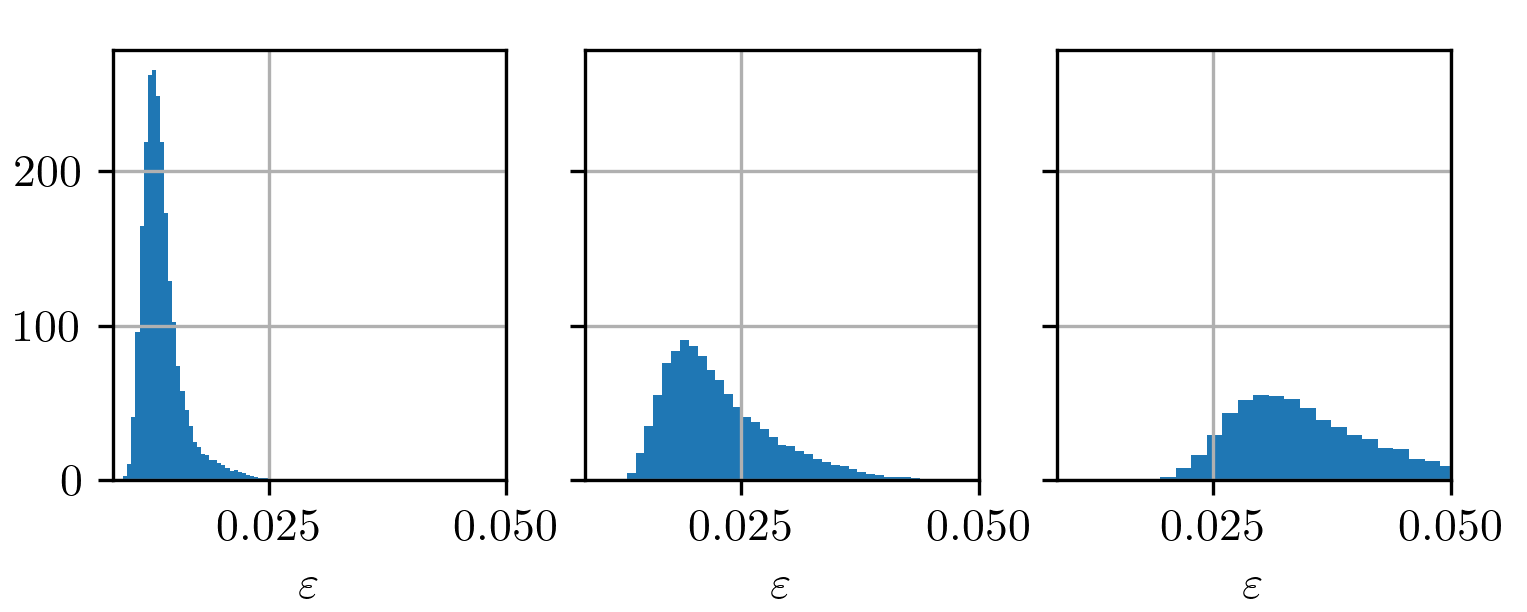}
    \caption{PDFs of the relative reconstruction error of IRMAE at $Re = 40, 100$ and $400$, from left to right.}
    \label{fig:irmae_errs}
\end{figure}

The probability density functions (PDFs) of the relative errors for each snapshot $\omega_i$ in the test datasets at $Re = \{40, 100, 400 \}$,
\begin{equation}
    \varepsilon_i = \frac{\| \omega_i - \mathscr{A}(\omega_i) \|}{\| \omega_i \|}\, ,
\end{equation}
is reported in figure \ref{fig:irmae_errs}.
Each $\omega_i$ in the test datasets is symmetry-augmented by random continuous and discrete symmetry operations.
As expected, the reconstruction accuracy decreases with $Re$ as the spatio-temporal complexity of the flow increases. 
Longer tails are also clear as $Re$ increases, likely reflecting the increasing proportion of high-dissipation bursting events in the test datasets.

%\appendix
% \begin{appen}


% \begin{acknowledgments}
% This work was supported by the Office of Naval Research (N XXXXXXXX) / Air Force Office of Scientific Research (AF XXXXXXXX). 
% 
% 
% \end{acknowledgments}

% \end{appen}

\bibliography{bib}

\end{document}